\documentclass[aps,preprint,showpacs,amsmath,superscriptaddress]{revtex4}

\usepackage{times}
\usepackage{hyperref}

\usepackage{graphicx}
\usepackage{dcolumn}
\usepackage{bm}

\begin{document}

\title{
Precision radiative corrections to the Dalitz plot of baryon semileptonic decays including the spin-momentum correlation of
the decaying baryon and the emitted charged lepton
}

\author{
M.\ Neri, A.\ Mart{\'\i}nez
}
\affiliation{
Escuela Superior de F\'{\i}sica y Matem\'aticas del IPN, Apartado Postal 75-702, M\'exico, D.F.\ 07738, M\'exico
}

\author{
Rub\'en Flores-Mendieta
}

\affiliation{
Instituto de F{\'\i}sica, Universidad Aut\'onoma de San Luis Potos{\'\i}, \'Alvaro Obreg\'on 64, Zona Centro,
San Luis Potos{\'\i}, S.L.P.\ 78000, M\'exico
}

\author{
J.\ J.\ Torres
}

\affiliation{
Escuela Superior de C\'omputo del IPN, Apartado Postal 75-702, M\'exico, D.F. 07738, M\'exico
}

\author{
A.\ Garc{\'\i}a
}

\affiliation{
Departamento de F{\'\i}sica, Centro de Investigaci\'on y de Estudios Avanzados del IPN, Apartado Postal 14-740,
M\'exico, D.F.\ 07000, M\'exico
}

\date{\today}

\begin{abstract}
We calculate the radiative corrections to the angular correlation between the polarization of the decaying baryon and the
direction of the emitted charged lepton in the semileptonic decays of spin one-half baryons to order $(\alpha/\pi)(q/M_1)$.
The final results are presented, first, with the triple integration of the bremsstrahlung photon ready to be performed
numerically and, second, in an analytical form. The results are useful in the analysis of the Dalitz plot of precision
experiments involving light and heavy quarks and is not compromised to fixing the form factors at predetermined values. It
is assumed that the real photons are kinematically discriminated. Otherwise, our results have a general model-independent
applicability.
\end{abstract}


\maketitle

In this paper we want to obtain precision radiative corrections (RC) to the ${\hat {\mathbf s}_1} \cdot {\hat {\mathbf l}}$
correlation of baryon semileptonic decays (BSD), $A\to Bl\overline{\nu}_l$ (where $A$ and $B$ are spin one-half baryons, $l$
is the charged lepton, and $\nu_l$ is the accompanying antineutrino or neutrino as the case may be, ${\hat {\mathbf s}_1}$ is
the polarization of $A$ and ${\hat {\mathbf l}}$ is the momentum direction of $l$) to order $(\alpha/\pi)(q/M_1)$, where $q$
is the momentum transfer and $M_1$ is the mass of $A$. However, it is not possible to obtain the results for this
correlation by a simple change of variables in the final results for other observables. One is forced to return to the early
steps of the analysis and perform there the appropriate change of variables. A program must be followed \cite{rfm02}. To this
order of approximation, the model independence of RC is ensured following the approach of Refs.~\cite{sirlin,low}. We
will assume that even if real photons go undetected, they can be kinematically discriminated so that we will limit our
calculations to the three-body region of BSD \cite{rfm02}.

The calculation we envisage here is very long and tedious. However, a substantial amount of effort can be saved by keeping a
close parallelism with previous calculations. In a recent work \cite{jjt04} we obtained the RC to the above order of
precision for the ${\hat {\mathbf s}_1} \cdot {\hat {\mathbf p}_2}$ correlation (where ${\hat {\mathbf p}_2}$ is the direction
of $B$). The analysis of this Ref.~\cite{jjt04} can be applied here step by step and if we use the same conventions and
notation the final results for the ${\hat {\mathbf s}_1} \cdot {\hat {\mathbf l}}$ correlation are in exact parallelism with
the final results for the ${\hat {\mathbf s}_1} \cdot {\hat {\mathbf p}_2}$ correlation of this reference.

This allows us to present directly the compact form of the final results of the RC to the Dalitz plot with the
${\hat {\mathbf s}_1} \cdot {\hat {\mathbf l}}$ correlation explicitly exhibited, namely,
\begin{equation}
d\Gamma_i = d\Omega \left[ A_0^\prime+\frac{\alpha}{\pi} \Theta_{iI} - {\hat {\mathbf s}_1} \cdot {\hat {\mathbf l}} \left(
A_0^{\prime\prime}+\frac{\alpha}{\pi} \Theta_{iII} \right) \right],  \label{eq:rdcf}
\end{equation}
where the index $i=C,N$ corresponds to charged decaying baryon (CDB) and the neutral decaying baryon (NDB) cases,
respectively. Here $\Theta_{iI} = B_1^\prime (\Phi_i+I_{i0}) + B_{i1}^{\prime\prime}\Phi_i^\prime+C_i^\prime$ and
$\Theta_{iII} = B_2^\prime (\Phi_i+I_{i0}) + B_{i2}^{\prime\prime}\Phi_i^\prime+C_i^{(s)}$.

In Eq.~(\ref{eq:rdcf}) both the virtual and the bremsstrahlung RC are included. The first two terms within the curly brackets
correspond to the unpolarized Dalitz plot. We shall only display the new results here. The explicit form of all other terms
can be readily identified using Ref.~\cite{jjt04}. Thus, $B_2^\prime = Ep_2y_0 \tilde{Q}_6 + El\tilde{Q}_7$,
$B_{C2}^{\prime\prime} = Ep_2y_0 Q_8+ ElQ_9$, and $B_{N2}^{\prime\prime} = M_1p_2y_0 Q_{N8} + M_1lQ_{N9}$. The other new
contribution $C_C^{(s)}$ of the polarized terms in $\Theta_{CII}$ consists of the sum of three terms, namely,
$C_{C}^{(s)}=p_2l/(2\pi) \sum_{R=I}^{III}\int_{-1}^{y_0}dy \int_{-1}^1dx\int_0^{2\pi}d\varphi_k
|M_R|^2/D$, where
\begin{eqnarray*}
\left| M_I\right|^2 & = & \frac{\beta^2 (1-x^2)}{2(1-\beta x)^2} \frac{E}{M_1} \left\{ -\frac{M_1D}{l}\tilde{Q}_6+M_1xQ_9 +
\frac{p_2y}{E}\left[-D+E(1+\beta x)\right]Q_{10}\right. \\
&   & \mbox{} + \left[ (p_2y+l)(1-\beta x) + \beta(E+E_\nu^0+p_2xy -D)\right] Q_{11} \\
&   & \mbox{} + \left[ \frac{p_2y+l}{E}D-l(1-\beta x) + \beta(E+E_\nu^0-p_2xy) \right] Q_{12}
+ lQ_{13} \\
&   & \mbox{} + \left.\left[ -\frac{p_2y+l}{E}D - l(1-\beta x) + \beta(E+E_\nu^0+p_2xy) \right] Q_{14}\right\},
\end{eqnarray*}
\begin{eqnarray*}
\left|M_{II}\right|^2 & = & \left. \frac{1}{2(1-\beta x)}\right\{ p_2y \tilde Q_6 + l \tilde Q_7 + R_1p_2y Q_8
+ \left\{R_1l+[R_2 E_\nu+w(R_1+1)]x\right\}Q_9 \\
&   & \mbox{} -\frac{p_2y}{M_1} \left[\frac{D\omega}{E}+ R_2({\mathbf p}_2 \cdot {\hat {\mathbf k}}+lx)
-2(1+R_1)\omega \right] Q_{10}  + \frac{l}{M_1} wQ_{13} - R_2 E_\nu x Q_{15} \\
&   & \mbox{} + \frac{p_2l}{M_1} \left\{\frac{2\omega}{p_2}(1-\beta x) - {\hat {\mathbf p}_2} \cdot {\hat {\mathbf k}}
\left[R_2+\frac{\omega}{E} \right] - R_2 \frac{x}{l}[{\hat {\mathbf p}_2} \cdot( {\mathbf k} - {\mathbf p}_\nu) -  ly]
+ \frac{\omega}{E} x y\right\}Q_{11} \\
&   & \mbox{} + \frac{p_2}{M_1} \left\{ R_2{\hat {\mathbf p}_2} \cdot[ {\hat {\mathbf k}}(p_2y+l+\omega x)
- {\mathbf p}_\nu x] + \beta\omega \left[\frac{D(p_2y+2l)}{p_2l}-
{\hat {\mathbf p}_2} \cdot {\hat {\mathbf k}}-xy\right] \right\}Q_{12} \\
&   & \mbox{} + \left. \frac{p_2}{M_1} \left[ R_2[-{\mathbf p}_2 \cdot {\hat {\mathbf k}}y + x(p_2+ly)]
-{\hat {\mathbf p}_2} \cdot {\hat {\mathbf k}}(R_2l+\beta\omega) + \frac{\omega y}{E}(lx-D) \right] Q_{14}
\right\},
\end{eqnarray*}
and
\begin{eqnarray*}
\left|M_{III}\right|^2 & = & \left. \frac{1}{1-\beta x} \frac{l}{M_1}\right\{ 2E_\nu (x^2-1)Q_{16}
+ (E_\nu-\beta {\hat {\mathbf l}} \cdot {\mathbf p}_\nu)(1-x^2)Q_{17} \\
&   & \mbox{} + \beta (p_2y+l+\omega x)(1-x^2) Q_{18} + (-E_\nu + D + {\hat {\mathbf l}} \cdot {\mathbf p}_\nu x)Q_{19} \\
&   & \mbox{} + (E_\nu - D - {\hat {\mathbf l}} \cdot {\mathbf p}_\nu x)Q_{20}
+ \beta [x(D-E_\nu) + {\hat {\mathbf l}} \cdot {\mathbf p}_\nu]Q_{21} \\
&   & \mbox{} + \frac{\omega x}{2l} (D-E_\nu+\beta {\hat {\mathbf l}} \cdot {\mathbf p}_\nu)Q_{22}
+ \frac{\omega}{2l} [\beta (D-2E_\nu)+(E_\nu+\beta {\hat {\mathbf l}} \cdot {\mathbf p}_\nu)x]Q_{23} \\
&   & \mbox{} + \left. \frac{E_\nu \omega}{2l} (1-\beta x)xQ_{24} - \frac{\omega}{2l} (p_2y+l+\omega x)(1-\beta x)Q_{25}
\right\}.
\end{eqnarray*}

The NDB case contains $C_N^{(s)}=C_C^{(s)}+C_{NA}^{(s)}$, where $C_{NA}^{(s)}=\tilde{C}_{I}^{(s)}+\tilde{C}_{II}^{(s)}+
\tilde{C}_{III}^{(s)}$. The explicit forms of these latter are $\tilde{C}_{i}^{(s)}=D_3 \rho_i+D_4\rho_i^\prime$,
$i=I,II,III$, with $D_3=2(-g_1^2+f_1g_1)$ and $D_4=2(g_1^2+f_1g_1)$, and
\begin{equation}
\rho_I = \frac{p_2^2l^2\beta }{\pi M_1}\int_{-1}^{y_0}dy \int_{-1}^1dx\int_0^{2\pi}\frac{d\varphi_k}{D(1-\beta x)}
(-y+x{\hat {\mathbf k}} \cdot {\hat {\mathbf p}_2}), \label{eq:ro1}
\end{equation}
\begin{equation}
\rho_I^\prime = \frac{p_2^2l}{\pi M_1}\int_{-1}^{y_0}dy \int_{-1}^1dx\int_0^{2\pi}\frac{d\varphi_k}{D(1-\beta x)}
(-y+x{\hat {\mathbf k}} \cdot {\hat {\mathbf p}_2})(-D+lx), \label{eq:ro1prima}
\end{equation}
\begin{equation}
\rho_{II} = \frac{p_2^2l}{4\pi M_1}\int_{-1}^{y_0}dy\int_{-1}^1dx \int_0^{2\pi}\frac{d\varphi_k}{D}E_\nu
\left[ y+\frac{\beta y-{\hat {\mathbf k}} \cdot {\hat {\mathbf p}_2}} {1-\beta x} x\right], \label{eq:ro2}
\end{equation}
\begin{equation}
\rho_{II}^\prime = \frac{p_2^2l}{4\pi M_1}\int_{-1}^{y_0}dy \int_{-1}^1dx\int_0^{2\pi}\frac{d\varphi_k}{D}\left[
{\hat {\mathbf k}} \cdot {\hat {\mathbf p}_2} +\frac{\beta y- {\hat {\mathbf k}} \cdot {\hat {\mathbf p}_2}}
{1-\beta x}\right] {\mathbf p}_\nu \cdot {\hat {\mathbf l}}, \label{eq:ro2prima}
\end{equation}
\begin{eqnarray}
\rho_{III} & = & \frac{p_2l}{2\pi M_1}\int_{-1}^{y_0}dy\int_{-1}^1dx \int_0^{2\pi} \left. \frac{d\varphi_k}{D(1-\beta x)}
\right\{ \left[ l(x-\beta)-2\omega(1-\beta x)\right] E_\nu x \nonumber \\
&   & \mbox{} - \left. l\left[ E_\nu(1-\beta x) -\beta {\mathbf p}_\nu \cdot (x{\hat {\mathbf k}} - {\hat {\mathbf l}})
\right] \right\}, \label{eq:ro3}
\end{eqnarray}
\begin{eqnarray}
\rho_{III}^\prime & = & \frac{p_2l}{2\pi M_1}\int_{-1}^{y_0}dy \int_{-1}^1dx\int_0^{2\pi}\frac{d\varphi_k}{D(1-\beta x)}
\left\{ -[\beta l (1-x^2) + 2\omega (1-\beta x)] {\mathbf p}_\nu \cdot {\hat {\mathbf l}} \right. \nonumber \\
&   & \mbox{} - \left. l(E_\nu-{\mathbf p}_\nu \cdot {\hat {\mathbf k}}) + (-{\mathbf l} \cdot {\mathbf p}_\nu
+E_\nu lx) x\right\}. \label{eq:ro3prima}
\end{eqnarray}

The above expressions contain triple integrals that can be integrated numerically. This is our first final result and covers
both the CDB and NDB cases.

These integrals can be performed analytically. A convenient rearrangement is
$C_C^{(s)} = \sum_{i = 1}^9 Q_{i+5}\Lambda_i + \sum_{i = 6}^{15} Q_i\Lambda_{i+4} + \sum_{i = 16}^{25} Q_i\Lambda_{i+4}$.
The form factors appear in the $Q_{i}$ functions, which are all collected in Appendix A of Ref.~\cite{jjt04}. The triple
integrals are contained in the $\Lambda_i$. In organizing the first summand of $C_C^{(s)}$ with one running index $i$
it was necessary to introduce $\Lambda_2=\Lambda_3=0$, because $Q_7$ and $Q_8$ do not appear in this equation. Skipping
details, we get
\begin{equation*}
\Lambda_1 = -Ep_2\theta_0, \qquad \quad \Lambda_4 = \frac{E^2p_2}{2}(Y_2-Y_3),
\end{equation*}
\begin{equation*}
\Lambda_5 = -\frac{Ep_2}{2M_1}[ -2Z_1+\beta lp_2Y_1+Z_2+\beta p_2(y_0-1)\theta_0],
\end{equation*}
\begin{equation*}
\Lambda_6 = \frac{Ep_2}{2M_1} [l^2Y_3-2\beta l \theta_0+\beta l(E+E_\nu^0)Y_2+Z_1],
\end{equation*}
\begin{equation*}
\Lambda_7 = \frac{p_2l}{2M_1}\left\{ p_2(y_0-1)\theta_0+l [ 2\theta_0-EY_3+(E+E_\nu^0) Y_2] + p_2lY_1
+ \frac{Z_2-Z_1}{\beta} \right\},
\end{equation*}
\begin{equation*}
\Lambda_8 = \frac{Ep_2l^2}{2M_1}Y_2,
\end{equation*}
\begin{equation*}
\Lambda_9 = -\frac{p_2l}{2M_1} \left\{p_2(y_0-1) \theta_0 + l(2\theta_0+EY_3- (E+E_\nu^0) Y_2)
+ \frac{Z_2-Z_1}{\beta} + p_2lY_1) \right\},
\end{equation*}
\begin{equation*}
\Lambda_{10} = \frac{p_2}{2}\zeta_{11}, \qquad \quad
\Lambda_{11} = \frac{p_2l^2}{2}\theta_3,
\end{equation*}
\begin{equation*}
\Lambda_{12} = -\frac{p_2}{2}\left[ \frac{Z_3}{E}+\zeta_{10} \right],
\end{equation*}
\begin{eqnarray*}
\Lambda_{13} & = & \left. \frac{Ep_2}{2}\right\{ (E_\nu^0+2E-l\beta)(1-\beta^2)\theta_2 - [(1-\beta^2)(E_\nu^0+3E)
+ E] \theta_3 + 2E(1-\beta^2)\theta_4 \\
&   & \mbox{} - \left. 3l\theta_5 - (1-\beta^2)\theta_6 + (5E-E_\nu^0-l\beta)
\frac{\theta_7}{2E} + \frac{\theta_9}{2E} + l\beta \theta_{10}-\frac{\beta}{2}\theta_{14}-3\eta_0\right\},
\end{eqnarray*}
\begin{equation*}
\Lambda_{14} = -\frac{p_2}{2M_1}\left[ \frac{p_2l}{2}(1-y_0)(\theta_0-2\eta_0) - \frac{E_\nu^0}{E}Z_3+(1-\beta^2)
\zeta_{22}-\frac{\zeta_{31}}{2E}-2\zeta_{21}\right],
\end{equation*}
\begin{eqnarray*}
\Lambda_{15} & = & -\frac{p_2}{2M_1}\left\{\frac{p_2l}{2}\eta_0(y_0-1) + l(1-\beta^2)\chi_{12} - l\chi_{11} - 2l^2\eta_0
+ p_2^2(\eta_0-\gamma_0+l\beta \theta_3) \right. \\
&   & \mbox{} - \left. \frac{\zeta_{21}}{2} + \frac{X_3-X_4}{l}+\frac{\chi_{30}}{2l}\right\},
\end{eqnarray*}
\begin{eqnarray*}
\Lambda_{16} & = & -\frac{p_2}{2M_1}\left\{ \frac{p_2l(y_0-1)}{2}(\theta_0-\eta_0) + \frac{E_\nu^0}{E}Z_3-\beta p_2l
(2ly_0+p_2)\theta_3 - \beta X_2 - \frac{(X_3-X_4)}{l} \right. \\
&   & \mbox{} + \left. 2l\beta \zeta_{11}+\frac{\zeta_{21}}{2}+\frac{\beta}{2}\chi_{21}
- \frac{\chi_{30}}{2l}+p_2^2(\gamma_0-\eta_0) \right\},
\end{eqnarray*}
\begin{equation*}
\Lambda_{17} = \frac{p_2l^2}{2M_1}\left[ \frac{\theta_7}{2}-E(\theta_3-\theta_4) \right],
\end{equation*}
\begin{eqnarray*}
\Lambda_{18} & = & -\frac{p_2}{2M_1}\left\{ \frac{p_2l(1-y_0)}{2}(\eta_0+\theta_0) -E[(1-\beta^2) \zeta_{12} - \zeta_{11}] - 2EZ_2 \right. \\
&   & \mbox{} + \left. l [(1-\beta^2) \chi_{12}-\chi_{11}] + p_2^2(\gamma_0-\eta_0-\beta l\theta_3)
- \frac{E+E_\nu^0}{E}Z_3 \right\},
\end{eqnarray*}
\begin{eqnarray*}
\Lambda_{19} & = & -\frac{Ep_2}{2}\left\{ E_\nu^0\left(-\frac{\gamma_0}{E}+\beta^2\theta_3\right) - \frac{\eta_0}{E}
\left( E-\frac{E_\nu^0}{2}\right) - (1-\beta^2) \left[ \frac{\theta_6}{2}-E(\theta_2-\theta_3) \right] \right. \\
&   & \mbox{} + \left. (2-\beta^2) \left[ \frac{\theta_7}{2}-E(\theta_3-\theta_4)\right] + \frac{1}{4E}
\left[\theta_9-2l^2\theta_3+6E^2(\theta_3-\theta_4-\beta \theta_5) \right] \right\},
\end{eqnarray*}
\begin{equation*}
\Lambda_{20} = \frac{2p_2E^2}{M_1}\left\{ \eta_0-(1-\beta^2)\left[ \frac{\theta_7}{2}-E_\nu^0\theta_3-E (\theta_3-\theta_4)
\right] - E_\nu^0 (\theta_4+\beta \theta_5) + \beta \left(\frac{\theta_{14}}{2}-l\theta_{10}\right)
\right\},
\end{equation*}
\begin{eqnarray*}
\Lambda_{21} & = & \frac{p_2E^2}{M_1}\left\{ -(1-\beta^2) \left( \frac{\zeta_{11}}{E}+(E_\nu^0+l\beta)\theta_3\right)
+ (E_\nu^0+\beta l)(\theta_4+\beta \theta_5) + \frac{\zeta_{10}}{E}\right. \\
&   & \mbox{} + \left. \beta^2 \left(-\eta_0 + p_2Y_1 + \frac{\theta_{21}}{2}
- l\theta_{20}\right) \right\},
\end{eqnarray*}
\begin{eqnarray*}
\Lambda_{22} & = & \frac{Ep_2l}{M_1} \left\{ \frac{\zeta_{10}}{l} + \beta p_2Y_1 + lY_3 + \frac{\beta^2-1}{\beta}
\left[ \frac{\zeta_{11}}{E} + \frac{\theta_7}{2} - E(\theta_3-\theta_4) - \eta_0\right] \right. \\
&   & \mbox{} + \left. \left[ \frac{\theta_{14}}{2} - l\theta_{10} + \frac{\beta}{2}(\theta_{21}-2l\theta_{20})
\right] \right\},
\end{eqnarray*}
\begin{eqnarray*}
\Lambda_{23} & = & \frac{p_2E^2}{M_1}\left\{\beta^2[E\theta_4-(E+E_\nu^0)\theta_3+\theta_0] + (1+2\beta^2)\eta_0
- (1-\beta^2) \left[ \frac{\theta_7}{2}-E\left(\theta_3-\theta_4\right) \right] \right. \\
&   & \mbox{} + \left. \frac{\zeta_{10}-\zeta_{11}}{E} + \beta \left( \frac{\theta_{14}}{2}-l\theta_{10}\right) \right\} ,
\end{eqnarray*}
\begin{eqnarray*}
\Lambda_{24} & = & \frac{p_2E^2}{M_1}\left\{ \beta^2[(E+E_\nu^0)\theta_3 - E\theta_4 -\theta_0]
+ \frac{\zeta_{11}-\zeta_{10}}{E} - (2\beta^2+1)\eta_0 \right. \\
&   & \mbox{} + \left. (1-\beta^2) \left[ \frac{\theta_7}{2} - E(\theta_3-\theta_4) \right]
- \frac{\beta}{2} \theta_{14} + \beta^2 E\theta_{10}\right\},
\end{eqnarray*}
\begin{equation*}
\Lambda_{25} = \frac{p_2l^2}{M_1}\left[E_\nu^0\theta_4+\theta_0-(E_\nu^0+\beta l) \theta_3-\frac{\zeta_{11}}{E}\right],
\end{equation*}
\begin{eqnarray*}
\Lambda_{26} & = & \frac{Ep_2}{4M_1}\left\{-\frac{\zeta_{21}}{E} - 2(\zeta_{11}-\zeta_{10}) - (E_\nu^0+\beta l)
\left[ \theta_7- 2E(\theta_3-\theta_4) - 2\eta_0\right] \right. \\
&   & \mbox{} + \left. 2p_2ly_0(\theta_3-\theta_4) - \beta \eta_0 \left[ l-\frac{p_2}{2}(y_0-1) \right] \right\},
\end{eqnarray*}
\begin{eqnarray*}
\Lambda_{27} & = & \left. \frac{p_2}{4M_1}\right\{ 2\beta l(p_2ly_0\theta_3-\zeta_{11}) - E(2E_\nu^0\beta^2-E_\nu^0+l\beta)
[\theta_7-2E(\theta_3-\theta_4)] + \zeta_{20} - \zeta_{21} \\
&   & \mbox{} + \left. l^2\eta_0 - \frac{p_2l(y_0^2-1)}{2} + E(\beta^2-1)[\theta_9-2l^2\theta_3+6E^2
(\theta_3-\theta_4-\beta \theta_5)] \right\} ,
\end{eqnarray*}
\begin{equation*}
\Lambda_{28} = \frac{p_2l}{4M_1}\left[ E_\nu^0(\theta_{14}-2l\theta_{10}) + l\eta_0+\frac{p_2\eta_0}{2} (y_0-1) \right\},
\end{equation*}
\begin{equation*}
\Lambda_{29} = -\frac{p_2l\eta_0}{8M_1} [ 2l+p_2(y_0-1)].
\end{equation*}

As was mentioned earlier we are using the same notation as in Ref.~\cite{jjt04}. There should arise no confusion, the list of
$\Lambda_i$ in this reference applies only to the ${\hat {\mathbf s}_1} \cdot {\hat {\mathbf p}_2}$ correlation and the above
list applies only to the ${\hat {\mathbf s}_1} \cdot {\hat {\mathbf l}}$ correlation. However, the $\theta_i$, $\gamma_0$,
$\eta_0$, $\zeta_{pq}$, $\chi_{mn}$, and the functions $X_i$, $Y_i$, and $Z_i$ given in Appendix B of that reference are
common to both sets of $\Lambda_i$.

The $\rho_i$ functions, after performing the integrations, become
\begin{equation*}
\rho_I = \frac{2p_2l}{M_1}(\chi_{11}-\chi_{10}-\beta \zeta_{11}),
\end{equation*}
\begin{eqnarray*}
\rho_I^\prime & = & \frac{2p_2l}{M_1}\left[ \frac{p_2(y_0-1)}{2\beta^2}[2\beta^2\eta_0+(\beta^2-1) \theta_0]
+ \frac{\zeta_{10}-\zeta_{11}}{\beta} \right. \\
&   & \mbox{} + \left. \frac{\chi_{11}-\chi_{10}}{\beta^2} + EE_\nu^0\theta_5 + El\theta_{10}\right],
\end{eqnarray*}
\begin{equation*}
\rho_{II} = \frac{p_2}{2M_1}\left[ \frac{E_\nu^0}{\beta}(\beta \zeta_{11}-\chi_{11}+\chi_{10})
- \frac{\zeta_{21}}{2} + \frac{\chi_{21}-\chi_{20}}{2\beta}\right],
\end{equation*}
\begin{eqnarray*}
\rho_{II}^\prime & = & -\frac{p_2}{2M_1}\left[ \frac{p_2l(1-y_0)}{2}(\theta_0+\eta_0) - p_2l^2(Y_1-Y_5)
- (E+E_\nu^0)(\zeta_{10}-\zeta_{11}) \right. \\
&   & \mbox{} + \left. \beta p_2^2l{\cal I} - l^2(\theta_0-\beta p_2\theta_{12})
+ \frac{\chi_{20}-\chi_{21}}{2\beta} + \frac{\zeta_{21}}{2} \right],
\end{eqnarray*}
\begin{eqnarray*}
\rho_{III} & = & \frac{p_2l}{M_1}\left[ \frac{m^2}{l}\left( \eta_0-\frac{\theta_7}{2}
+ (E+E_\nu^0)(\theta_3-\theta_4) \right) + \frac12 (E-2E_\nu^0)(\theta_{14}-2l\theta_{10}) \right. \\
&   & \mbox{} - \left. EE_\nu^0\theta_5 - \frac12 p_2(y_0^2-1) + \beta p_2l\theta_{12}
-l(\theta_0 + E_\nu^0\theta_4-EY_3) \right],
\end{eqnarray*}
\begin{eqnarray*}
\rho_{III}^\prime & = & \frac{p_2l}{M_1}\left[ \beta \zeta_{11} + p_2lY_1 + ElY_3 - \frac{m^2}{2l}\theta_7
- EE_\nu^0\theta_5 + \frac{E}{2}(\theta_{14}-2l\theta_{10}) - l\theta_0 \right. \\
&   & \mbox{} + \left. \frac{l}{2}(\theta_{21}-2l\theta_{20}) + \frac{E}{\beta}(E+E_\nu^0)(\theta_3-\theta_4)
+ \eta_0 \left(\frac{E}{\beta} - 2l + \frac{p_2}{2}(y_0-1)\right) \right].
\end{eqnarray*}

As in the case of $\Lambda_i$, the use of the same notation for the $\rho_i$ functions here and in Ref.~\cite{jjt04} should
not lead to confusions. Again both these sets use the same algebraic expressions on their right-hand sides. They can all be
found in Appendix B of Ref.~\cite{jjt04}. All the detailed definitions are found there.

Our results for ${\hat {\mathbf s}_1} \cdot {\hat {\mathbf l}}$ are general within our approximations. We covered the CDB
and NDB cases. They can be used in the other four charge assignments of baryons involving heavy quarks and whether the
charged lepton is $e^\pm$, $\mu^\pm$, $\tau^\pm$. They are model independent and are not compromised to fixing the form
factors at prescribed values. They are given in two forms. The first one is the triple numerical integration form, in which
the integrations over the real photon variables are explicitly exhibited and may be performed numerically. In the second one
all those integrations were calculated analytically. Our calculations rely heavily on previous results and we have provided
the references to the many expressions calculated before. Let us point out that there is a third way of using our RC results
in the form of numerical arrays, as discussed in Ref.\cite{jjt04}, which may be the more practical in experimental analyses.
We shall not discuss this further, due to lack of space. The results will be useful in high statistics experiments of
hyperon semileptonic decays (involving only the light quarks $u,d,s$) and of medium statistics experiments of heavy quark
baryons (involving the transitions $c\rightarrow s,d$ and $b\rightarrow c,u,d$).

The authors are grateful to CONACYT (Mexico) for partial support. J.J.T., A.M., and M.N.\ were partially supported by
COFAA-IPN. R.F.M.\ was also partially supported by FAI-UASLP. He also wishes to express his gratitude to Mauro Piccini for
useful discussions.

\end{document}